\documentclass[preprint,authoryear,12pt]{elsarticle}
\usepackage{amsmath,amsfonts,amssymb}
\usepackage{theorem}
\usepackage{graphicx}
\usepackage{color}
\usepackage{longtable}
\usepackage{natbib}
\newcommand{\half}{\frac{1}{2}}
\newcommand{\Half}{\frac{3}{2}}

\newcommand{\be}{\begin{equation}}
\newcommand{\ee}{\end{equation}}
\newcommand{\bea}{\begin{eqnarray}}
\newcommand{\eea}{\end{eqnarray}}

 \journal{BioSystems}

\begin{document}

\begin{frontmatter}

\pagestyle{empty}

 \title{{\bf A Minimum Principle in \\ 
Codon-Anticodon  Interaction}}

 \author[label1,label2]{A. Sciarrino\corref{cor1}}
 \ead{sciarrino@na.infn.it}
\author[label3]{P. Sorba}

\address[label1]{ Dipartimento di Scienze Fisiche, Universit{\`a} di Napoli ``Federico II'' \\   Complesso Universitario di Monte S. Angelo \\ Via Cinthia, I-80126 Napoli, Italy }

\address[label2]{I.N.F.N., Sezione di Napol \\ sciarrino@na.infn.it\\~\\~}

\address[label3]{Laboratoire d'Annecy-le-Vieux de Physique Th{\'e}orique LAPTH\\
CNRS, UMR 5108, associ{\'e}e {\`a} l'Universit{\'e} de Savoie \\
BP 110, F-74941 Annecy-le-Vieux Cedex, France\\sorba@lapp.in2p3.fr}

\ead{sorba@lapp.in2p3.fr }

\cortext[cor1]{\texttt{Corresponding author: Tel +39/081676807 - Fax +39/081676346} }

\begin{abstract}
    Imposing a minimum principle in the framework of the so called crystal basis model of the genetic code, we determine the structure of the minimum set  of  anticodons which allows the translational-transcription for animal mitochondrial code. The results are in very good agreement with the observed anticodons.  
\end{abstract}

\begin{keyword} 
genetic code \sep codon \sep anticodon  \\
 
\rightline{Report no: DSF-012/11 - LAPTH-040/11}
 
\end{keyword}

\end{frontmatter}

\newpage
\pagestyle{empty}
\phantom{blankpage}
\newpage
\pagestyle{plain}
\setcounter{page}{1}
\baselineskip=16pt

\section{Introduction}

The translational-transcription process from DNA to proteins is  a very complex process carried on in several steps. A key step is the translation from coding sequences of nucleotides in mRNA to the proteins chaines. In this process a role is played by the tRNA in which a triplet of nucleotides (anticodon) pairs to the triplet of nucleotides (codon) reading the genetic information.
Since there are 60 codons  (in mitochondrial code) specifying amino acids, the cell should contain 60 different tRNA molecules, each with a different anticodon in order to have a pairing codon anticodon following the usual Watson-Crick pattern, i.d. the pairing respectively  between the nucleotides C and G, and U and A. Actually, however, the number of observed anticodons is less than 60. This implies that an anticodon  may pair to  more than one codon. Already in the middle of the sixties,
 it was realized that the pairing anticodon-codon does not follow the standard rule and Crick \citep{Crick} proposed, on the basis of  the base-pair stereochemistry, the  ``wobble hypothesis".   According to this hypothesis a single tRNA type,  with a a specified anticodon, is able to recognize two or more codons in particular differing only in the third nucleotide, i.e only the first two nucleotides of a codon triplet  in mRNA have the standard precise pairing with the bases of the tRNA anticodon while the first nucleotide in the anticodon may pair to more than a nucleotide in the third position of the codon.  

This rule has been subsequently widely confirmed and extended, with  a better understanding of the chemical nucleotide modifications, for a review see \citep{Agris}. 
Since the years seventies the questions  were raised  \citep{Jukes}: how many anticodons do we need?  which anticodons do manifest? 

In order to explain which anticodon do manifest two main hypothesis have been advanced: \begin{enumerate}
\item The conventional wobble versatility hypothesis assumes that the the first position of anticodon should have G (U) to read for codon with Y (respectively R) in third position.
\item The codon adaptation hypothesis states that the first position of anticodon should pair the most abondant codon in the family of synonymous codons.
\end{enumerate}
For a comparison and discussion of the two hypothesis in fungal mitochondrial genomes  and for marine bivalve mitochondrial genomes, see \citep{CX} and \citep{YL}.

 In order to have a correct translation process between codons and amino-acids in the mitochondrial code we need a minimum number of 22 anticodons. In fact, in this code, the 20 amino-acids (a.a) are encoded by  2 sextets,
6 quadruplets and 12 doublets of codons. Considering a sextet as the sum of a quadruplet and a doublet, we need to dispose at least  of  22 anti-codons, of which 8 should  ``read" the quadruplets and 14 the doublets.  Indeed this seems to happen for the mitochondria of animals  \citep{Sprinzl,HJJR,WN,Nial,NikW}.  
The data seem to confirm the empirical  rule that the most used anticodons have as second and third nucleotide, respectively, the complementary  to the first and second nucleotide of the codons, while the first nucleotide is U for the anticodons pairing the quadruplets, G  and U for the anticodons pairing, respectively, the doublets ending with a pyrimidine and  with a purine, with exception of Met.

The aim of this paper is to propose a mathematical approach in the framework of the  ``crystal basis model" model  of the genetic code  \citep{FSS98},  to determine which anticodon is chosen to translate the genetic information stored into the quadruplets and the doublets of codons. More generally, the idea is to require
the minimization of a suitable operator or function, mathematically expressed in terms of the quantities defined in the model, to explain why and which anticodon is used to   ``read"  more than a codon\footnote{We do not discuss here the chemical modified structure of the nucleotides, e.g see \citep{Agris}.}.

Let us very quickly recall the main ideas of the model introduced in \citep{FSS98}, for a review and some applications see \citep{FSS01}. In that paper we have proposed a mathematical framework in 
which the codons appear as composite states of nucleotides.  The four 
nucleotides being assigned to the fundamental irreducible representation (irrep.) of 
the quantum group ${\cal U}_{q}(su(2) \oplus su(2))$ in the limit $q \to 
0$, the codons are obtained as tensor product of nucleotides.  Indeed, the 
properties of quantum group representations in the limit $q \to 0$, or 
crystal basis, are crucial to take into account the fact that a codon is an ordered triple of nucleotides.
The nucleotide content of the $({\bf \half,\half)}$ (fundamental) representation of ${\cal 
U}_{q \to 0}(su(2) \oplus su(2))$, i.e. the eigenvalues of  $ J_{H,3}, \, J_{V,3}$, is chosen as follows:
\be 
	\mbox{C} \equiv (+\half,+\half) \qquad \mbox{U} \equiv (-\half,+\half) 
	\qquad \mbox{G} \equiv (+\half,-\half) \qquad \mbox{A} \equiv 
	(-\half,-\half)
	\label{eq:gc1}
\ee 
where the first $su(2)$ - denoted  $su(2)_{H}$- corresponds to the distinction between the purine bases A,G and the pyrimidine ones C,U and the second one - denoted $su(2)_{V}$ - corresponds to the complementarity rule C/G and U/A, 
Thus to represent a codon, we have to perform the tensor product of three 
$(\half,\half)$ or fundamental representations of ${\cal U}_{q \to 0}(su(2) \oplus 
su(2))$  and we get the results, reported in Table \ref{table-mitan}, where we have also written the observed anticodon for the mitochondria of animals taken from   \citep{Sprinzl}.
Really in the present paper we use the minimum principle in a reduced form as we are only interested to find the composition of the minimum number of anticodons. However in the last section we hint at some more general application of our schema. 

The paper is organised  as follows: in Sec. 2 we present the minimum principle, in Sec. 3 we apply the principle to the mitochondrial code for animals and we compare our theoretical results with the data of \citep{Sprinzl}. In the final Section we give a short summary and  some highlights on future developments and applications. 

 \section{The  ``minimum" principle}
 
Given a  codon\footnote{In the paper we use the notation $N = C, A, G, U.; \; \;  R =   G, A. \;(purine)  ;\;\; Y =   C, U. \; (pyrimidine)$.} $XYZ$ ($X,Y,Z \in \{ C, A, G, U\}$) we conjecture that an anticodon    
 $\,X^aY^aZ^a$, where $\,Y^aZ^a=Y_cX_c$,  $N_c$ denoting  the nucleotide complementary to the nucleotide $N$ according to the Watson-Crick pairing rule\footnote{This property is observed to be verified in most, but not in all, the observed cases. To simplify we shall assume it.},
pairs to the codon $XYZ$, i.e. it is most used to ``read" the codon $XYZ$ if it minimizes the operator ${\cal T}$,  explicitly written in eq.(\ref{eq:T}) and computed between the 
 ``states", which can be read from Table \ref{table-mitan}, describing the codon and anticodon in the  ``crystal basis model". We write both codons (c) and anticodons (a) in $5" \to 3"$ direction. As an anticodon is antiparallel to codon, the 1st nucleotide (respectively the 3rd nucleotide) of the anticodon is paired to the 3rd (respectively the 1st) nucleotide of the codon.
  
\be
{\cal T} = 8 c_H \,\vec{ J_H^c} \cdot  \vec{J_H^a} + 8  c_V \, \vec{J_V^c} \cdot \vec{ J_V ^a} \label{eq:T}
\ee
where:
\begin{itemize}

  \item $c_H.  c_V$ are constants depending on the  ``biological species" and weakly depending on the encoded a.a., as we  will later specify.
 
\item $J_H^c ,  J_V^c$  (resp. $J_H^a ,  J_V^a$) are the labels of  $U_{q \to 0}(su(2)_H \oplus su(2)_V)$  specifying the state \citep{FSS98} describing the codon $XYZ $ (resp. the anticodon $NY_cX_c$ pairing the codon  $XYZ$).

\item  $\vec{ J_{\alpha}^c} \cdot  \vec{J_{\alpha}^a}$ ($\alpha = H, V$) should be read as 
\be
\vec{ J_{\alpha}^c} \cdot  \vec{J_{\alpha}^a} =
\frac{1}{2}\left\{\left(\vec{J_{\alpha}^c}  \oplus \vec{J_{\alpha}}^a \right)^2 - ( \vec{J_{\alpha}^c})^2 -  ( \vec{J_{\alpha}^a})^2 \right\}
\ee
and   $\vec{J_{\alpha}^c}  \oplus  \vec{J_{\alpha}^a} \equiv \vec{J_{\alpha}^T}$  stands for the irreducible representation which the codon-anticodon state under consideration belongs to, the tensor product of   $\vec{J_{\alpha}^c}$ and $\vec{J_{\alpha}^a}$ being performed according to the rule of \citep{Kashi}, choosing the codon as first vector and the anticodon as second vector.  Note that 
$ \vec{J_{\alpha}}^2$ should be read as the Casimir operator whose eigenvalues are given by $
J_{\alpha}(J_{\alpha} +1 )$.
\end{itemize}

For example the value  of ${\cal T}$  between the anticodon UUU and the codon AAC
is, using Table \ref{coef-d}:
\be
 < UUU| {\cal T}|AAC> =  -6 \, c_H +  18 \, c_V
 \ee
As we are interested in finding the composition of the 22  anticodons, minimun number to ensure a faithful translation,  we shall assume that the used anticodon for each quartet and each doublet is the one 
which minimizes the  averaged value of the operator  given in eq.(\ref{eq:T}), the average being performed over the 4 (2) codons  for quadruplets (doublets), see next  section.
Indeed it is well known that synonymous codons are not used with equal frequency.  Therefore, in finding the structure of the anticodons in the minimum set, it appears reasonable that the codon usage probability plays a role in the determination of the chosen anticodon. 
If the codon $XYZ$ is used more frequently than the codon $XYZ'$, the codon $XYZ$ should give an impact larger than the codon $XYZ'$ in determining the choice of the anticodon in the ${\cal T}$.  Therefore, ${\cal T}_{av}$ 
appears  more appropriate than ${\cal T}$ for our purpose.

\section{Structure of the minimum number of anticodons}

According to our conjecture on the existence of a minimum principle  we  determine,  for each quadruplet (q) and each doublet (d),  the anticodon which minimizes the averaged value  ${\cal T}_{av}$ of the operator ${\cal T}$ (see below).  We analyse separately the case of quadruplets and doublets.
 
\subsection{Quadruplets}

Let us give an example of what we mean by averaged value of ${\cal T}$. For example let us 
consider the anticodon  CAC for the a.a. Val,  we have to compute
\bea
&&  {\cal T}_{av}(CAC, Val) = \sum _{N} \; P^q_N \, <CAC|{\cal T}|GUN>  \nonumber \\
&& =  P^q_C  \, <CAC|{\cal T}|GUC> +  P^q_U  \, <CAC|{\cal T}|GUU>   \nonumber \\
&& + P^q_G  \,  <CAC|{\cal T}|GUG>+ P^q_A  \,   <CAC|{\cal T}|GUA>  \nonumber  \\
&& = 2 (P^q_C + P^q_U + P^q_G +P^q_A) \, c_H + (6 P^q_C + 6 P^q_U + 2 P^q_G + 2P^q_A) \, c_V\nonumber \\
&& = 2 \, c_H  + [6 P^q_Y + 2(1- P^q_Y)]\, c_V
\eea
In the computation we have to take into account the codon usage frequency or relative percentage of the appearance of each codon in the quadruplet and we have denoted with $P^q_N$ the codon usage frequency for codon ending with N.
Really  we need to introduce the following four positive frequencies
 $P_Y^q$ $P_R^q$, $P_S^q$,$P_W^q$, with the normalization condition:
\be
P_Y^q + P_R^q = P_S^q + P_W^q = 1
\ee
where, respectively, $P_Y^q$, $P_R^q$, $P_S^q$ and $P_W^q$ denote the relative usage frequency of the codons ending  with nucleotides C,U (pyrimidine),  G,A (purine),  C,G and  U,A.
From Table  \ref{coef-q} we can compute the value which we report in Table \ref{avT-q}.
 
  \subsection{Doublets}

In the computation we have to take into account the codon usage frequency in the doublet. Now we need to introduce the following four positive frequencies
 $P_C^d$, $P_U^d$, $P_G^d$, $P_A^d$, with the normalization condition
\be
P_C^d + P_U^d = P_G^d + P_A^d = 1
\ee
 As example let us compute the averaged value of ${\cal T}$ for Asp. 
Let us consider the anticodon  CUC  we have to compute
\bea
&&{\cal T}_{av}(CUC, Asp) = \sum _{Y} \;  P_Y^d\,  <CUC|{\cal T}|GAY>  = P_C^d\, <CUC|{\cal T}|GAC> + P_U^d\,  <CUC|{\cal T}|GAU>    \nonumber  \\
&& = 2 \, c_H +18 \, c_V\
\eea
From Table  \ref{coef-d},  we can compute the values, which we report in Table \ref{avT-d2}.

 Let us remark that:
 \begin{itemize}
\item  for all a.a. the contribution of $su_V(2)$ verifies the same property than for the quadruplets and, moreover, is not depending on the codon usage;

\item for 4  a.a.  the contribution of $su_V(2)$ is the same for all anticodon.

 \end{itemize}
 
 From the above remarks we easily realize that the case of doublets is more complicated than the one of the quadruplets. In some sense the contribution of $su_V(2)$  plays a role only in establishing the  most preferred anticodon. Moreover,  as we do not want a priori to exclude any anticodon, we have to face the possibility that an anticodon can be chosen to read for more than one doublet. In order to avoid this problem, in contradiction with the requirement of a faithful translation process, we make the following choice:
 \begin{enumerate}
 \item the sign of the constant $c_H$ for the doublets ending with a purine is the opposite of the sign of
 the doublets ending with a pyrimidine with the same dinucleotide (if it does exist)\footnote{We call dinucleotide the first two nucleotids of the codon.}. 
 \item the sign of  $c_H$ for the 8 weak dinucleotides encoding doublets is positive for the following 4 doublets UUY, UAY, AUY, AAY and negative for the remaining 4, i.e. CAY, UGY, AGY, GAY.
 \end{enumerate} 
  and fix the following procedure, while considering doublets with the same
dinucleotide:

 \begin{enumerate}
\item first we select, among the four possible anticodons, the one giving the lowest
value for  ${\cal T}$ averaged on the two codons of each doublet and assign this anticodon to
the corresponding doublet.

\item then the anticodon reading the second doublet is chosen between the two ones
containing as a first nucleotide a purine, resp. a pyrimidine, if the first
nucleotide of the anticodon already determined for the first doublet is a
pyrimidine (resp. a purine).
\end{enumerate}  
 
As an illustration, we take the case of the Cys and Trp amino acids. The anticodon
GCA can minimize both of them, but more Cys (due to the value $-6$ for the $c_V$ coefficient) than Trp
(with value $2$ for the same  $c_V$ coefficient). Thus GCA will be taken as the anticodon
relative to Cys, while the choice for the Trp anticodon will be made between UCA and
CCA, that is the two candidates with a pyrimidine as a first nucleotide, the
anticodon GCA starting with a purine.

Let us remark that, even if the above assumptions seem rather ad hoc,  indeed a general symmetric pattern shows up: for half of a.a. $c_H$ is positive and for the other half is negative;  the first set of 4 dinucleotides involves only  `weak" nucleotides, the second one a  ``strong" nucleotide; the dinucleotides XY and YX correspond to the same sign.

 \subsection{Discussion}

  For all quadruplets, from Table  \ref{coef-q}, we remark that for $c_H > 0$ and  $c_V < 0$ the  anticodon  minimizing  the average value of ${\cal T}$ has the composition $ UX_cY_c$. For Leu, Val and Thr a very weak condition for codon usage frequency has to be satisfied, i.e for the first two a.a. $  P^q_S  > 0,25$ and for the last one $  P^q_Y  > 0,125$\footnote{ The constraint on the codon usage frequency  can be released, imposing a suitable condition between $c_H$ and  $c_V $.}.
   The results are in agreement with the observed anticodons, see \citep{Sprinzl} and Table \ref{table-mitan}.
 
 For doublets, we remark that, with the choice of sign of  $c_H $  above specified and  $c_V > 0$ for all a.a., the  anticodons  minimizing  the average value of ${\cal T}$  are  in agreement with the observed anticodon, see \citep{Sprinzl} and Table \ref{table-mitan}.
 We summarize in Table \ref{min-d} the results for the doublets.

\begin{table}[htbp]
\begin{tabular}{|c||c||c||c|} \hline 
a.a & sign $ c_{H}$ & anticodon & note   \\
\hline \hline
His &  - & GUG & $P^d_C > 0,25$ \\
\hline  
Gln & + &  UUG &  $P^d_G > 0,25$   \\
\hline\hline
Phe & - &   GAA  &  \\
\hline  
Leu & + &   UAA &  \\
\hline \hline
Cys &+ &   GCA &   \\
 \hline  
Trp & - & UCA &   \\
\hline \hline
Tyr &  - & GUA & \\
\hline  \hline
Ser & + &  GCU &  \\
\hline  \hline 
Asp & + & GUC &  $P^d_C > 0,25$ \\
\hline  
Glu &- & UUC &    $P^d_G > 0,25$ \\
\hline \hline
Ile &  + & GAU  & \\
\hline  
Met  & - &  CAU &   \\
\hline\hline
Asn & - &   GUU &  \\
\hline  
Lys & + &   UUU &  \\
\hline \hline
\end{tabular}
\centering \caption{Anticodon minimizing  the operator ${\cal T}$, averaged over the two codons,  for any amino acid encoded by a doublet, specifying the sign of $c_H$. } 
 \label{min-d}
 \end{table}
   Let us remark that we find that for Met the anticodon is not UAU, as it should be expected from the empirical rule  above quoted, but CAU which seems in agrement with the data, see \citep{Sprinzl}.
 
\section{Conclusions}

We have found that the anticodons minimizing the conjectured operator ${\cal T}$ given in eq.(\ref{eq:T}), averaged  over the concerned multiplets, are in very good agreement, the results depending only on the signs of the two coupling constants,
with the observed ones, even if we have made comparison with a limited database.

The fact that the crystal basis model is able to explain, in a  relatively simple way,  the observed anticodon-codon pairing which has its roots on the stereochemical properties of nucleotides \citep{LimCur} strongly suggests that our modeling is able to incorporate some crucial features of the complex physico-chemical structure of the genetic code.

It is rather clear that the operator ${\cal T}$ can be looked at as a codon-anticodon interaction operator. In this spirit, given a codon, the selected anti-codon for fixing the corresponding
amino-acid, is determined as the one which minimizes the interaction.
  It is hard at this point to be more specific on the
physico-chemical aspects of this quantity. 

Indeed, the analysis in \citep{PO}  suggests that the use of the wobble behaviour is  also dictated from the requirement of the optimization of  translational efficiency and  
in \citep{Lehmann} the free-energy change of anticodon-codon interaction has been put into relation with the dissociation time, depending on the  molecular structure of the encoded amino-acid, while in \citep{LimCur} emphasis has been put in the hydrogen and ionic bondings.

It is intriguing that complex behaviour involving thermodynamical considerations as well as evolutionary effects can be  cast in a single simple mininum principle.

It might be interesting to note, among the different previously obtained
applications of the crystal basis model,
that this model has previously allowed to establish  a pattern of
correlations between the physico-chemical properties
  of the amino-acids and the assignment of the corresponding coding codons in the model \citep{FSS02}. Incidentally let us remark that the model  explains the symmetry codon anticodon remarked in \citep{WN}.  Let us stress that our modeling has
a very peculiar feature which makes it very different from the standard 4-letter alphabet, used to identify the nucleotides, as well as with the usual modeling  of nucleotide chain as spin chain.  Indeed the identification of the nucleotides with the fundamental irrep. of ${\cal U}_{q}(su(2)_H \oplus su(2)_V)$ introduces a sort of  double ``bio-spin", which allows the description of any ordered sequence of $n$ nucleotides as as state of an irrep. and allows to describe interactions using the standard powerful mathematical language used in physical spin models.

In the present paper we have faced the problem to find the structure of the mimimum set of anticodons and, then, we have used a very simple form for the operator ${\cal T}$, with the main aim to present a simple, mathematical modeling of the extremely complex codon-anticodon interaction.
 We have not at all  discussed the possible appearance of any  other anticodon, which should require a more quantitative discussion.
For such analysis, as well as
for the eukaryotic code, the situation may be different and more than an anticodon may pair to  a quartet.  
For this future aim we have here reported Tables \ref{coef-q} and  \ref{coef-d}. 

The pattern, which in the general case may show up, is undoubtedly more complicated, depending on the biological species and on the concerned biosynthesis process, but it is natural to argue that the usage of anticodons  exhibits the general feature to assure an ``efficient"  translation process by  a number of anticodons, minimum with respect to the involved constraints.
 A more refined and quantitative analysis, as well as comparison with other organisms, which should require more data,  depends on  the value of these constants.
Very likely, it might happen that the assumption of the  ``universal" feature of $c_H$ and $c_V$ should be released and that  the expression of the operator eq.(\ref{eq:T}) should be modified, for example by adding a term of  ``spin-spin" interaction of the type
\be
4 g_H \, J_{H,3}^c J_{H,3}^a +   4 g_V \, J_{V,3}^c J_{V,3}^a 
\ee
where the values of  $ J_{H,3}$ and $ J_{V,3} $,  both for codons and anticodons, can be read out  from their nucleotide composition, see  Table \ref{table-mitan}.
However the pattern which shows up in Tables \ref{coef-q} and \ref{coef-d}, with the values of the coefficients  equal in pairs, strongly suggests that the minimum number of anticodons should be 32 (3 for the sextets, 2 for quadruplets and triplet and 1 for doublets and singlets).

\begin{table}[htbp]
\footnotesize
\begin{center}
\begin{tabular}{|cc|cc|rr|c||cc|cc|rr|c||}
\hline
codon & a.a. & $J_{H}$ & $J_{V}$ & $J_{3,H}$ & $J_{3,V}$& anticodon & codon & 
a.a. &   $J_{H}$ & $J_{V}$ & $J_{3,H}$ & $J_{3,V}$ & anticodon\\
\hline
\Big. CCC & P & $\Half$ & $\Half$ & $\Half$ & $\Half$ & & UCC & S & 
$\Half$ & $\Half$ & $\half$ & $\Half$ & \\
\Big. CCU & P & $(\half$ & $\Half)^1$ & $\half$ & $\Half$ &  
&UCU & 
S & $(\half$ & $\Half)^1$ & $-\half$ & $\Half$&  \\
\Big. CCG & P & $(\Half$ & $\half)^1$ & $\Half$ & $\half$ &{\LARGE \color{red} \bf UGG} & UCG & 
S & $(\Half$ & $\half)^1$ & $\half$ & $\half$  & {\LARGE \color{red} \bf   \bf UGA} \\
\Big. CCA & P & $(\half$ & $\half)^1$ & $\half$ & $\half$  & &UCA & 
S & $(\half$ & $\half)^1$ & $-\half$ & $\half$ & \\[1mm]
\hline
\Big. CUC & L & $(\half$ & $\Half)^2$ & $\half$ & $\Half$ & & UUC & 
F & $\Half$ & $\Half$ & $-\half$ & $\Half$ & \\
\Big. CUU & L & $(\half$ & $\Half)^2$ & $-\half$ & $\Half$ & 
& UUU & 
F & $\Half$ & $\Half$ & $-\Half$ & $\Half$ &  {\LARGE \color{blue} \sl GAA}\\
\Big. CUG & L & $(\half$ & $\half)^3$ & $\half$ & $\half$ & {\LARGE \color{red} \bf UAG} & UUG & 
L & $(\Half$ & $\half)^1$ & $-\half$ & $\half$ &   \\
\Big. CUA & L & $(\half$ & $\half)^3$ & $-\half$ & $\half$ & &UUA & 
L & $(\Half$ & $\half)^1$ & $-\Half$ & $\half$ & {\LARGE \color{blue} \sl UAA} \\[1mm]
\hline
\Big. CGC & R & $(\Half$ & $\half)^2$ & $\Half$ & $\half$ & &UGC & 
C & $(\Half$ & $\half)^2$ & $\half$ & $\half$& \\
\Big. CGU & R & $(\half$ & $\half)^2$ & $\half$ & $\half$ &   
& UGU & 
C & 
$(\half$ & $\half)^2$ & $-\half$ & $\half$  & {\LARGE \color{blue} \sl  GCA}  \\
\Big. CGG & R & $(\Half$ & $\half)^2$ & $\Half$ & $-\half$ &  {\LARGE \color{red} \bf UCG} & UGG & 
W & $(\Half$ & $\half)^2$ & $\half$ & $-\half$ &\\
\Big. CGA & R & $(\half$ & $\half)^2$ & $\half$ & $-\half$ & &UGA & 
W & $(\half$ & $\half)^2$ & $-\half$ & $-\half$ & {\LARGE \color{blue} \sl  UCA} \\[1mm]
\hline
\Big. CAC & H & $(\half$ & $\half)^4$ & $\half$ & $\half$ &  &UAC & 
Y & $(\Half$ & $\half)^2$ & $-\half$ & $\half$ &  \\
\Big. CAU & H & $(\half$ & $\half)^4$ & $-\half$ & $\half$ & {\LARGE \color{blue} \sl GUG} & UAU & 
Y & $(\Half$ & $\half)^2$ & $-\Half$ & $\half$ &  {\LARGE \color{blue} \sl GUA} \\
\Big. CAG & Q & $(\half$ & $\half)^4$ & $\half$ & $-\half$ &&UAG & 
 {\bf Te}r & $(\Half$ & $\half)^2$ & $-\half$ & $-\half$ & ----- \\
\Big. CAA & Q & $(\half$ & $\half)^4$ & $-\half$ & $-\half$ & {\LARGE \color{blue} \sl UUG}  &UAA & 
 {\bf Ter} & $(\Half$ & $\half)^2$ & $-\Half$ & $-\half$& ----- \\[1mm]
\hline
\Big. GCC & A & $\Half$ & $\Half$ & $\Half$ & $\half$ & & ACC & T & 
$\Half$ & $\Half$ & $\half$ & $\half$ &  \\
\Big. GCU & A & $(\half$ & $\Half)^1$ & $\half$ & $\half$ && ACU & 
T & $(\half$ & $\Half)^1$ & $-\half$ & $\half$ & \\
\Big. GCG & A & $(\Half$ & $\half)^1$ & $\Half$ & $-\half$ &  {\LARGE \color{red} \bf UGC} & ACG & 
T & $(\Half$ & $\half)^1$ & $\half$ & $-\half$& {\LARGE \color{red} \bf UGU} \\
\Big. GCA & A & $(\half$ & $\half)^1$ & $\half$ & $-\half$ & &ACA & 
T & $(\half$ & $\half)^1$ & $-\half$ & $-\half$ &\\[1mm]
\hline
\Big. GUC & V & $(\half$ & $\Half)^2$ & $\half$ & $\half$ &  & AUC & 
I & $\Half$ & $\Half$ & $-\half$ & $\half$& \\
\Big. GUU & V & $(\half$ & $\Half)^2$ & $-\half$ & $\half$ & & AUU & 
I & $\Half$ & $\Half$ & $-\Half$ & $\half$&   {\LARGE \color{blue} \sl GAU} \\
\Big. GUG & V & $(\half$ & $\half)^3$ & $\half$ & $-\half$ & {\LARGE \color{red} \bf UAC}& AUG & 
M & $(\Half$ & $\half)^1$ & $-\half$ & $-\half$&  \\
\Big. GUA & V & $(\half$ & $\half)^3$ & $-\half$ & $-\half$ & &
AUA &  M & $(\Half$ & $\half)^1$ & $-\Half$ & $-\half$ &  {\LARGE \color{blue} \sl CAU}\\[1mm]
\hline
\Big. GGC & G & $\Half$ & $\Half$ & $\Half$ & $-\half$ & & AGC & S 
& $\Half$ & $\Half$ & $\half$ & $-\half$ & \\
\Big. GGU & G & $(\half$ & $\Half)^1$ & $\half$ & $-\half$ &&AGU & 
S & $(\half$ & $\Half)^1$ & $-\half$ & $-\half$ &  {\LARGE \color{blue} \sl GCU}\\
\Big. GGG & G & $\Half$ & $\Half$ & $\Half$ & $-\Half$ & {\LARGE \color{red} \bf UCC} &AGG & 
{\bf Ter} & $\Half$ & $\Half$ & $\half$ & $-\Half$&----- \\
\Big. GGA & G & $(\half$ & $\Half)^1$ & $\half$ & $-\Half$ & 
&AGA & {\bf Ter} & $(\half$ & $\Half)^1$ & $-\half$ & $-\Half$ & ----- \\[1mm]
\hline
\Big. GAC & D & $(\half$ & $\Half)^2$ & $\half$ & $-\half$ && AAC & 
N & $\Half$ & $\Half$ & $-\half$ & $-\half$  & \\
\Big. GAU & D & $(\half$ & $\Half)^2$ & $-\half$ & $-\half$ & {\LARGE \color{blue} \sl GUC} & AAU & 
N & $\Half$ & $\Half$ & $-\Half$ & $-\half$ & {\LARGE \color{blue} \sl  GUU} \\
\Big. GAG & E & $(\half$ & $\Half)^2$ & $\half$ & $-\Half$ &   & AAG & 
K & $\Half$ & $\Half$ & $-\half$ & $-\Half$ & \\
\Big. GAA & E & $(\half$ & $\Half)^2$ & $-\half$ & $-\Half$ &{\LARGE \color{blue} \sl UUC} & AAA & 
K & $\Half$ & $\Half$ & $-\Half$ & $-\Half$& {\LARGE \color{blue} \sl UUU} \\[1mm]
\hline
\end{tabular}
\caption{
The vertebral mitochondrial code. The upper label denotes 
different irreducible representations. 
 We list the most used anticodons for mitochondria of animals, see \citep{Sprinzl}. In bold-red (italic-blue) the anticodons reading quadruplets (resp. doublets).}
\label{table-mitan}
\end{center}
\end{table}

\clearpage
\newpage
\begin{table}[htbp]
\begin{tabular}{|c|c||c|c||c|c||c|c||c|c||} \hline 
a.a & codon &$ K_{H,C}$ & $ K_{V,C}$   &$ K_{H,U}$ & $ K_{V,U}$ &$ K_{H,G}$ & $ K_{V,G}$&   $ K_{H,A}$ & $ K_{V,A}$  \\
\hline \hline
Pro & CCC & 18 & -10& -6 & -10& 18 & -30& -6 & -30 \\
& CCU & 6 & -10& -10 & -10& 6 & -30& -10 & -30 \\
& CCG & 18 & -6& -6 & -6& 18 & -10& -6 & -10 \\
& CCA & 6 & -6 & -10 & -6 & 6 & -10 & -10 & -10 \\
\hline 
Leu & CUC & 2 & -10& -10 & -10& 2 & -30& -10 & -30 \\
& CUU & 2 & -10& 6 & -10& 2 & -30& 6 & -30 \\
& CUG & 2 & -6& -10 & -6& 2 & -10& -10 & -10 \\
& CUA & 2 & -6 & 6 & -6 & 2 & -10& 6 & -10 \\
\hline
Arg & CGC & 18 & 2 & -6 & 2 & 18 & -6 & -6 & -6 \\
& CGU & 6 & 2 & -10 & 2 & 6 & -6 & -10 & -6 \\
& CGG & 18 & 2 & -6 & 2 & 18 & 2 & -6 & 2 \\
& CGA & 6 & 2 & -10 & 2 & 6 & 2 & -10 & 2 \\
\hline
Ala  & GCC & 18 & 6 & -6 & 6 & 18 & -22 & -6 & -22 \\
& GCU & 6 & 6 & -10 & 6 & 6 & -22 & -10 & -22 \\
& GCG & 18 & 2 & -6 & 2 & 18 & 6 & -6 & 6 \\
& GCA & 6 & 2 & -10 & 2 & 6 & 6 & -10  & 6 \\
\hline
Gly  & GGC & 18 & 18 & -6 & 18 & 18 & -6 & -6 & -6 \\
& GGU & 6 & 18 & -10 & 18 & 6 & -6 & -10 & -6 \\
& GGG & 18 & 18 & -6 & 18 & 18 & 18 & -6 & 18 \\
& GGA & 6 & 18 & -10 & 18 & 6 & 18 & -10  & 18 \\
\hline 
Val & GUC & 2 & 6 & -10 & 6 & 2 & -22& -10 & -22 \\
& GUU & 2 & 6 & 6 & 6 & 2 & -22& 6 & -22 \\
& GUG & 2 & 2 & -10 & 2 & 2 & 6 & -10 & 6 \\
& GUA & 2 & 2 & 6 & 2 & 2 & 6 & 6 & 6 \\
\hline
Ser & UCC & 6 & -10& -10 & -10& 6 & -30& -10 & -30 \\
& UCU & 2 & -10& 2 & -10& 2 & -30& 2 & -30 \\
& UCG & 6 & -6& -10 & -6& 6 & -10& -10 & -10 \\
& UCA & 2 & -6 & 2 & -6 & 2 & -10 & 2 & -10 \\
\hline
Thr  & ACC & 6 & 6 & -10 & 6 & 6 & -22 & -10 & -22 \\
& ACU & 2 & 6 & 2 & 6 & 2 & -22 & 2 & -22 \\
& ACG & 6 & 2 & -10 & 2 & 6 & 6 & -10 & 6 \\
& ACA & 2 & 2 & 2 & 2 & 2 & 6 & 2 & 6 \\
\hline \hline
\end{tabular}
\centering \caption{Values of the coefficient multiplying $c_H$ ($K_H = 8\vec{ J_H^c} \cdot  \vec{J_H^a})$ and  $c_V $ $(K_V = 8\vec{ J_V^c} \cdot  \vec{J_V^a})$ computed  from the value of the tensor product of  the codon  $XYZ$ with the anticodon $NY_cX_c$,  for the quadruplets.} 
 \label{coef-q}
 \end{table}
 
 \clearpage
 
 \begin{table}[htbp]
\begin{tabular}{|c|c||c|c||c|c||c|c||c|c||} \hline 
a.a & codon &$ K_{H,C}$ & $ K_{V,C}$   &$ K_{H,U}$ & $ K_{V,U}$ &$ K_{H,G}$ & $ K_{V,G}$&   $ K_{H,A}$ & $ K_{V,A}$  \\
\hline \hline
His & CAC & 2 & 2 & -10 & 2 & 2 & -6 & -10 & -6 \\
& CAU & 2 & 2 & 6 & 2 & 2 & -6& 6 & -6 \\
\hline
Gln & CAG & 2 & 2 & -10 & 2 & 2 & 2 & 6 & 2 \\
& CAA & 2 & 2 & 6 & 2 & 2 & 2 & 6 & 2 \\
\hline
Phe & UUC & -10 & -10& -6 & -10& -10 & -30& -6 & -30 \\
& UUU & 6 & -10& 18 & -10& 6 & -30& 18 & -30 \\
\hline
Leu & UUG & -10 & -6& -6 & -6&-10 & -10& -6 & -10 \\
& UUA & 6 & -6 & 18 & -6 & 6 & -10 & 18 & -10 \\
\hline
Cys & UGC & 6 & 2 & -10 & 2 & 6 & -6 & -10 & -6 \\
& UGU & 2 & 2 & 2 & 2 & 2 & -6 & 2 & -6 \\
\hline
Trp & UGG & 6 & 2 & -10 & 2 & 6 & 2 & -10 & 2 \\
& UGA & 2 & 2 & 2 & 2 & 2 & 2 & 2 & 2 \\
\hline\hline
Tyr  & UAC & -10 & 2 & -6 & 2 & -10 & -6 & -6 & -6 \\
& UAU & 6 & 2 & 18 & 2 & 6 & -6 & 18 & -6 \\
\hline\hline
Asp & GAC & 2 & 18 & -10 & 18& 2 & -6& -10 & -6 \\
& GAU & 2 & 18& 6 & 18& 2 & -6& 6 & -6 \\
\hline
Glu & GAG & 2 & 18& -10 & 18& 2 & 18& -10 & 18 \\
& GAA & 2 & 18 & 6 & 18 & 2 & 18& 6 & 18 \\
\hline\hline
 Ile & AUC & -10 & 6& -6 & 6 & -10 & -22& -6 & -22 \\
& AUU & 6 & 6 & 18 & 6 & 6 & -22& 18 & -22 \\
\hline
Met & AUG & -10 & 2 & -6 & 2 &-10 & 6 & -6 & 6 \\
& AUA & 6 & 2 & 18 & 2 & 6 & 6 & 18 & 6 \\
 \hline \hline
Ser & AGC & 6 & 18 & -10 & 18 & 6 & -6 & -10 & -6 \\
& AGU & 2 & 18 & 2 & 18 & 2 & -6 & 2 & -6 \\
\hline\hline
Asn & AAC & -10 & 18 & -6 & 18& -10 & -6& -6 & -6 \\
& AAU & 6 & 18& 18 & 18 & 6 & -6& 18 & -6 \\
\hline
Lys & AAG & -10 & 18& -6 & 18& -10 & 18& -6 & 18 \\
& AAA & 6 & 18 & 18 & 18 & 6 & 18& 18 & 18 \\
 \hline \hline
\end{tabular}
\centering \caption{Values of the coefficient multiplying $c_H$ ($K_H = 8\vec{ J_H^c} \cdot  \vec{J_H^a})$ and  $c_V $ $(K_V = 8\vec{ J_V^c} \cdot  \vec{J_V^a})$ computed  from the value of the tensor product of  the codon  $XYZ$ with the anticodon $NY_cX_c$,  for the doublets.} 
 \label{coef-d}
 \end{table}
 
\clearpage

 \begin{table}[htbp]
\begin{tabular}{|c|c||c||c|} \hline 
a.a & anticodon & coeff. $ c_{H}$ &  coeff. $ c_{V}$  \\
\hline \hline
Pro & CGG &  18$P^q_S$ +6(1-$P^q_S$) & -10$P^q_Y$ - 6(1-$P^q_Y$)  \\
& UGG &  -6$P^q_S$ -10(1-$P^q_S$)& -10$P^q_Y$ - 6(1-$P^q_Y$) \\
& GGG &  18$P^q_S$ +6(1-$P^q_S$) & -30$P^q_Y$ - 10(1-$P^q_Y$) \\
& AGG &  -6$P^q_S$ -10(1-$P^q_S$& -30$P^q_Y$ - 10(1-$P^q_Y$)  \\
\hline 
Leu & CAG & 2 &   -10$P^q_Y$ - 6(1-$P^q_Y$)  \\
& UAG &   -10$P^q_S$ + 6(1-$P^q_S$) &  -10$P^q_Y$ - 6(1-$P^q_Y$)  \\
& GAG & 2 &  -30$P^q_Y$ - 10(1-$P^q_Y$) \\
& AAG &   -10$P^q_S$ + 6(1-$P^q_S$) &  -30$P^q_Y$ - 10(1-$P^q_Y$) \\
\hline
Arg & CCG & 18$P^q_S$ +6(1-$P^q_S$)  & 2   \\
& UCG & -6$P^q_S$ -10(1-$P^q_S$) & 2   \\
& GCG & 18$P^q_S$ +6(1-$P^q_S$) & -6$P^q_Y$ +2(1-$P^q_Y$)  \\
& ACG & -6$P^q_S$ - 10(1-$P^q_S$)&   -6$P^q_Y$ +2(1-$P^q_Y$)  \\
\hline
Ala  & CGC & 18$P^q_S$ + 6(1-$P^q_S$) &  6$P^q_Y$ + 2(1-$P^q_Y$)  \\
& UGC &  -6$P^q_S$ - 10(1-$P^q_S$) &  6$P^q_Y$ + 2(1-$P^q_Y$)   \\
& GGC &  18$P^q_S$ + 6(1-$P^q_S$) &  -22$P^q_Y$ + 6(1-$P^q_Y$)    \\
& AGC & -6$P^q_S$ - 10(1-$P^q_S$)  &   -22$P^q_Y$ + 6(1-$P^q_Y$)   \\
\hline
Gly  & CCC & 18$P^q_S$ + 6(1-$P^q_S$)  & 18  \\
& UCC &  -6$P^q_S$  -10(1-$P^q_S$)& 18  \\
& GCC & 18$P^q_S$ + 6(1-$P^q_S$) &   6$P^q_Y$  + 18(1-$P^q_Y$)  \\
& ACC &   -6$P^q_S$  -10(1-$P^q_S$) &  6$P^q_Y$  + 18(1-$P^q_Y$)  \\
\hline 
Val & CAC & 2 & 6$P^q_Y$  + 2(1-$P^q_Y$)  \\
&UAC & -10$P^q_S$ + 6(1-$P^q_S$) & 6$P^q_Y$  + 18(1-$P^q_Y$)   \\
& GAC & 2 &  -22$P^q_Y$ + 6(1-$P^q_Y$)    \\
& AAC & -10$P^q_S$ + 6(1-$P^q_S$) &  -22$P^q_Y$ + 6(1-$P^q_Y$)   \\
\hline
Ser & CGA & 6$P^q_S$ + 2(1-$P^q_S$) &  -10$P^q_Y$ - 6(1-$P^q_Y$) \\
& UGA &  -10$P^q_S$ + 2(1-$P^q_S$)  &   -10$P^q_Y$ - 6(1-$P^q_Y$) \\
& GGA &  6$P^q_S$ + 2(1-$P^q_S$)  &  -30$P^q_Y$ - 10(1-$P^q_Y$)\\
& AGA & -10$P^q_S$ + 2(1-$P^q_S$)  & -30$P^q_Y$ - 10(1-$P^q_Y$)    \\
\hline
Thr  & CGU &  6$P^q_S$ +2(1-$P^q_S$) & 6$P^q_Y$ + 2(1-$P^q_Y$)  \\
& UGU &  -10$P^q_S$ +2(1-$P^q_S$) & 6$P^q_Y$ + 2(1-$P^q_Y$)  \\
& GGU &   6$P^q_S$ +2(1-$P^q_S$)  & -22$P^q_Y$ + 6(1-$P^q_Y$)  \\
& AGU & -10$P^q_S$ +2(1-$P^q_S$)  & -22$P^q_Y$ + 6(1-$P^q_Y$)  \\
\hline \hline
\end{tabular}
\centering \caption{Value of the coefficients multiplying   $c_H $ and  $c_V $ in ${\cal T}_{av}$ , computed for any anticodon and averaged over the four codons for each quadruplet.} 
 \label{avT-q}
 \end{table}

\clearpage

\begin{longtable}{|c|c||c||c|} 
 
 \hline 
\multicolumn{1}{|c|} {\textbf a.a} &  \multicolumn{1} {|c|} {\textbf anticodon} & \multicolumn{1} {|c|}  {\textbf coeff. $ c_{H}$} &  
\multicolumn{1}{|c|} {\textbf  coeff. $ c_{V}$} \\
\hline \hline
\endfirsthead
\hline
\multicolumn {3}{l}  {\textbf Continued from previous page} \\ 
\hline
\multicolumn{1}{|c|} {\textbf a.a} &  \multicolumn{1} {|c|} {\textbf anticodon} & \multicolumn{1} {|c|}  {\textbf coeff. $ c_{H}$} &  
\multicolumn{1}{|c|} {\textbf  coeff. $ c_{V}$} \\
 \endhead
 
\multicolumn{2}{r}{\textbf Continued on next page} \\ \hline

\endfoot
\endlastfoot
His & CUG &  2 & 2  \\
& UUG &  -10$P^d_C$ + 6(1-$P^d_C$)& 2\\
& GUG & 2 & -6 \\
& AUG & -10$P^d_C$ + 6(1-$P^d_C$)& -6 \\
\hline  \hline
Gln & CUG &  2 & 2  \\
& UUG &  -10$P^d_G$ + 6(1-$P^d_G$)& 2\\
& GUG & 2 & 2 \\
& AUG &  6 & 2 \\
\hline\hline
Phe & CAA &  -10$P^d_C$ + 6(1-$P^d_C$) & -10  \\
& UAA &  -6$P^d_C$ + 18(1-$P^d_C$)& -10\\
& GAA & -10$P^d_C$ + 6(1-$P^d_C$) & -30 \\
& AAA & -6$P^d_C$ + 18(1-$P^d_C$)& -30 \\
\hline  \hline
Leu & CAA &  -10$P^d_G$ + 6(1-$P^d_G$) & -6  \\
& UAA &  -6$P^d_G$ + 18(1-$P^d_G$)& -6\\
& GAA & -10$P^d_G$ + 6(1-$P^d_G$) & -10 \\
& AAA & -6$P^d_G$ + 18(1-$P^d_G$)& -10 \\
\hline \hline
Cys & CCA &  6$P^d_C$ + 2(1-$P^d_C$) & 2  \\
& UCA &  -10$P^d_C$ + 2(1-$P^d_C$)& 2\\
& GCA & 6$P^d_C$ + 2(1-$P^d_C$) & -6 \\
& ACA & -10$P^d_C$ + 2(1-$P^d_C$)& -6 \\
\hline  \hline
Trp & CCA &  6$P^d_G$ + 2(1-$P^d_G$) & 2  \\
& UCA &  -10$P^d_G$ + 2(1-$P^d_G$)& 2\\
& GCA & 6$P^d_G$ + 2(1-$P^d_G$) & 2 \\
& ACA & -10$P^d_G$ + 2(1-$P^d_G$)& 2 \\
\hline \hline
Tyr & CUA &  -10$P^d_C$ + 6(1-$P^d_C$) & 2  \\
& UUA &  -6$P^d_C$ + 18(1-$P^d_C$)& 2\\
& GUA & -10$P^d_C$ + 6(1-$P^d_C$) & -6 \\
& AUA & -6$P^d_C$ + 18(1-$P^d_C$)& -6 \\
\hline  \hline
Ser & CCU &  6$P^d_C$ + 2(1-$P^d_C$) & 18  \\
& UCU &  -10$P^d_C$ + 2(1-$P^d_C$)& 18\\
& GCU & 6$P^d_C$ + 2(1-$P^d_C$) & -6 \\
& ACU & -10$P^d_C$ + 2(1-$P^d_C$)& -6 \\
\hline  \hline
Asp & CUC & 2 & 18  \\
& UUC &  -10$P^d_C$ + 6(1-$P^d_C$)& 18\\
& GUC &  2 & -6 \\
& AUC & -10$P^d_C$ + 6(1-$P^d_C$)& -6 \\
\hline  \hline
Glu & CUC &   2 & 18  \\
& UUC &  -10$P^d_G$ + 6(1-$P^d_G$)& 18\\
& GCA & 2 & 18 \\
& ACA & -10$P^d_G$ + 6(1-$P^d_G$)& 18 \\
\hline \hline
Ile & CAU &  -10$P^d_C$ + 6(1-$P^d_C$)& 6  \\
& UAU &  -6$P^d_C$ + 18(1-$P^d_C$)& 6\\
& GAU & -10$P^d_C$ + 6(1-$P^d_C$ & -22 \\
& AAU & -6$P^d_C$ + 18(1-$P^d_C$)& -22 \\
\hline  \hline
Met & CAU &   -10$P^d_G$ + 6(1-$P^d_G$)& 2  \\
& UAU &  -6$P^d_G$ +186(1-$P^d_G$)& 2\\
& GAU &  -10$P^d_G$ + 6(1-$P^d_G$)& 6 \\
& AAU &   18$P^d_G$ + 6(1-$P^d_G$) & 6 \\
\hline\hline
Asn & CUU &  -10$P^d_C$ + 6(1-$P^d_C$) & 18  \\
& UUU &  -6$P^d_C$ + 18(1-$P^d_C$)& 18\\
& GUU & -10$P^d_C$ + 6(1-$P^d_C$) & -6 \\
& AUU & -6$P^d_C$ + 18(1-$P^d_C$)& -6 \\
\hline  \hline
Lys & CUU &  -10$P^d_G$ + 6(1-$P^d_G$) & 18  \\
& UUU &  -6$P^d_G$ + 18(1-$P^d_G$)& 18\\
& GUU & -10$P^d_G$ + 6(1-$P^d_G$) & 18 \\ 
& AUU & -6$P^d_G$ + 18(1-$P^d_G$)& 18 \\
\hline \hline
\caption{ Value of the coefficients multiplying   $c_H $ and  $c_V $ in ${\cal T}_{av}$ , computed for any anticodon and averaged over the two codons for each doublet.} 
\label{avT-d2}
 \end{longtable}

\bibliographystyle{elsarticle-harv}
\setcitestyle{authoryear,round}

\end{document}